\documentclass[useAMS,usenatbib]{mn2e}

\usepackage[T1]{fontenc}
 \usepackage{pslatex}
\usepackage{amsmath}
\usepackage{amsfonts}
\usepackage{color}
\usepackage{url}
\usepackage{graphicx}
\usepackage{subfigure}
\usepackage{amssymb}
\usepackage{soul}
\usepackage{booktabs}
\usepackage{array}
\usepackage[utf8]{inputenc}
\inputencoding{latin1}
\inputencoding{utf8}

{\newif\ifnotend
\notendtrue
\def\veclist{ABCDEFGHIJKLMNOPQRSTUVWXYZabcdefghijklmnopqrstuvwxyz.}
\def\top#1#2.{#1}
\def\tail#1#2.{#2.}
\loop\expandafter\xdef\csname v\expandafter\top\veclist\endcsname%
{{\noexpand\bf\expandafter\top\veclist}}
\edef\veclist{\expandafter\tail\veclist}
\if\veclist.\notendfalse\fi\ifnotend\repeat}
\def\e{{\rm e}}

\def\E{{\cal E}}

\mathchardef\mhyphen="2D

\title[Orphan gamma-ray flares in blazars]{A fully-kinetic model for orphan gamma-ray flares in blazars}

\author[Sobacchi, N\"attil\"a, Sironi]{Emanuele Sobacchi$^{1}$\thanks{E-mail: es3808@columbia.edu}, Joonas N\"attil\"a$^{2,3}$, Lorenzo Sironi$^1$\\
$^1$ Department of Astronomy and Columbia Astrophysics Laboratory, Columbia University, 550 West 120th Street New York, NY 10027, USA\\
$^2$ Department of Physics and Columbia Astrophysics Laboratory, Columbia University, New York, NY 10027, USA\\
$^3$ Center for Computational Astrophysics, Flatiron Institute, 162 Fifth Avenue, New York, NY 10010, USA\\
}

\begin{document}

\date{}

\def\p{\partial}
\def\E{\textbf{E}}
\def\B{\textbf{B}}
\def\v{\textbf{v}}
\def\j{\textbf{j}}
\def\s{\textbf{s}}
\def\e{\textbf{e}}

\newcommand{\di}{\mathrm{d}}
\newcommand{\bfx}{\mathbf{x}}
\newcommand{\bfe}{\mathbf{e}}
\newcommand{\vlos}{\mathrm{v}_{\rm los}}
\newcommand{\Tspin}{T_{\rm s}}
\newcommand{\Tb}{T_{\rm b}}
\newcommand{\degree}{\ensuremath{^\circ}}
\newcommand{\Th}{T_{\rm h}}
\newcommand{\Tc}{T_{\rm c}}
\newcommand{\bfr}{\mathbf{r}}
\newcommand{\bfv}{\mathbf{v}}
\newcommand{\bfu}{\mathbf{u}}
\newcommand{\pc}{\,{\rm pc}}
\newcommand{\kpc}{\,{\rm kpc}}
\newcommand{\Myr}{\,{\rm Myr}}
\newcommand{\Gyr}{\,{\rm Gyr}}
\newcommand{\kms}{\,{\rm km\, s^{-1}}}
\newcommand{\de}[2]{\frac{\partial #1}{\partial {#2}}}
\newcommand{\cs}{c_{\rm s}}
\newcommand{\rb}{r_{\rm b}}
\newcommand{\rqu}{r_{\rm q}}
\newcommand{\bfOmega}{\pmb{\Omega}}
\newcommand{\bfOmegap}{\pmb{\Omega}_{\rm p}}
\newcommand{\bfXi}{\boldsymbol{\Xi}}

\maketitle

\begin{abstract}
Blazars emit a highly-variable non-thermal spectrum. It is usually assumed that the same non-thermal electrons are responsible for the IR-optical-UV emission (via synchrotron) and the gamma-ray emission (via inverse Compton). Hence, the light curves in the two bands should be correlated. Orphan gamma-ray flares (i.e., lacking a luminous low-frequency counterpart) challenge our theoretical understanding of blazars. By means of large-scale two-dimensional radiative particle-in-cell simulations, we show that orphan gamma-ray flares may be a self-consistent by-product of particle energization in turbulent magnetically-dominated pair plasmas. The energized particles produce the gamma-ray flare by inverse Compton scattering an external radiation field, while the synchrotron luminosity is heavily suppressed since the particles are accelerated nearly along the direction of the local magnetic field. The ratio of inverse Compton to synchrotron luminosity is sensitive to the initial strength of turbulent fluctuations (a larger degree of turbulent fluctuations weakens the anisotropy of the energized particles, thus increasing the synchrotron luminosity). Our results show that the anisotropy of the non-thermal particle population is key to modeling the blazar emission.
\end{abstract}

\begin{keywords}
galaxies: jets -- gamma-rays: galaxies -- radiation mechanisms: non-thermal -- plasmas -- turbulence
\end{keywords}


\section{Introduction}
\label{sec:introduction}

Blazars, Active Galactic Nuclei (AGN) launching a jet in the direction of the observer, are remarkably variable at all wavelengths. Variability at different wavelengths is usually correlated, but sometimes is not. A striking example of uncorrelated variability are the so-called ``orphan'' gamma-ray flares. During an orphan flare, the gamma-ray flux may increase by a factor of $10-100$ with respect to the quiescent level, while the IR-optical-UV flux does not show any significant variation \citep[e.g.][]{Krawczynski2004, Blazejowski2005, Hayashida2015, Macdonald2017}. In this paper we mostly focus on orphan GeV flares from Flat Spectrum Radio Quasars (FSRQ),\footnote{In FSRQ, gamma-rays are most likely emitted as a population of non-thermal electrons inverse Compton scatters off an external photon field, which may be produced by the Broad Line Region \citep[e.g.][]{Sikora1994, Sikora2009, GhiselliniTavecchio2009}. In BL Lacs, gamma-rays are instead emitted as the non-thermal electrons scatter the same synchrotron photons that they emit \citep[e.g.][]{Maraschi1992}.} which have been routinely reported since the launch of the {\it Fermi} Large Area Telescope \citep[e.g.][]{Macdonald2017}.

Orphan gamma-ray flares challenge the standard one-zone blazar emission model, which assumes the same electrons to emit the IR-optical-UV radiation via synchrotron, and the gamma-rays via inverse Compton (IC) \citep[e.g.][]{Maraschi1992, Sikora1994}. Hence, the IR-optical-UV and gamma-ray light curves should be well correlated.

Different explanations for the origin of orphan gamma-ray flares have been proposed \citep[for a review, see e.g.][]{Bottcher2019}, including (i) proton synchrotron emission \citep[e.g.][]{WeidingerSpanier2015}; (ii) the temporary enhancement of the external radiation field at the location of the emitting region \citep[e.g.][]{KusunoseTakahara2006, Macdonald2015, Tavani2015}; (iii) the injection of fresh non-thermal electrons into the emitting region, and the simultaneous decrease of the magnetic field \citep[e.g.][]{BottcherBaring2019, Lewis2019}. Though associating the flare to the injection of fresh non-thermal electrons may look attractive, the lack of a detectable synchrotron counterpart constrains the magnetic field to be well below the equipartition level. This requirement challenges the widely accepted paradigm that AGN jets are magnetically-dominated objects \citep[e.g.][]{BlandfordZnajek1977, Komissarov2007, Tchekhovskoy2011}.

The need for sub-equipartition magnetic fields can be tracked back to the textbook (and {\it ad hoc}) assumption that the distribution of the synchrotron-emitting particles is isotropic. Instead, we show that energetic particles move nearly along the direction of the local magnetic field. Then the radiated synchrotron power is heavily suppressed even if the magnetic field is large, while the IC power is practically the same as in the standard isotropic case.

In magnetically-dominated AGN jets, internal shocks are weak. Then viable non-thermal particle acceleration mechanisms are reconnection and turbulence. Since there is a huge separation of scales between the transverse size of the jet, which is the energy-carrying scale, and the particle Larmor radius, which is the scale where dissipation happens, we argue that dissipation proceeds through a turbulent cascade. Recently, fully-kinetic particle-in-cell (PIC) simulations have reached the maturity to simulate the energization of non-thermal particles in turbulent magnetically-dominated plasmas from first principles \citep[e.g.][]{Zhdankin+2017, Zhdankin+2018, zhdankin_20, ComissoSironi2018, ComissoSironi2019, Nattila2019, comisso_sobacchi_20}. Using highly-magnetized uncooled PIC simulations, \citet[][]{ComissoSironi2018, ComissoSironi2019} have shown that (i) particles are first energized due to strong non-ideal electric fields in large-scale reconnecting current sheets.\footnote{In reconnection-based models of blazar flares, it is usually assumed that the initial conditions can be modelled as a single large-scale Harris current sheet \citep[e.g.][]{Petropoulou+2016, Christie+2019}. Instead, in our model the properties of the large-scale current sheets are a self-consistent by-product of the turbulent plasma motions.} Since the reconnection electric field is nearly aligned with the local magnetic field, the distribution of the reconnection-accelerated particles is strongly anisotropic, with particles preferentially moving along the magnetic field;\footnote{In relativistic reconnection with weak guide fields, the reconnection electric field is nearly perpendicular to the magnetic field. Then reconnection-accelerated particles move in the direction perpendicular to the magnetic field (i.e., pitch angles are large). During the turbulent cascade, reconnection instead occurs in the regime of strong guide field, and the reconnection electric field is directed along the guide field. Then reconnection-accelerated particles move nearly along the direction of the guide field (i.e., pitch angles are small).} (ii) particles are further accelerated by scattering off turbulent magnetic fluctuations. The distribution of the scattered particles becomes increasingly isotropic at higher energies.

Comparing the gamma-ray luminosity with the extended emission from the radio lobes suggests that blazars are extremely radiative efficient \citep[e.g.][]{Nemmen2012}. Then turbulence should develop in the fast cooling regime, i.e. reconnection-accelerated particles should radiate most of their energy within the light crossing time of the system (otherwise, adiabatic losses become dominant and the jet is radiatively inefficient). Here we use large-scale two-dimensional PIC simulations to study radiative turbulence in the fast cooling regime. We show that (i) the local anisotropy of the reconnection-accelerated particles may suppress the synchrotron luminosity by orders of magnitude with respect to the IC luminosity; (ii) further energization due to scattering is inhibited since it operates on a longer timescale than cooling \citep[see also][]{NattilaBeloborodov2020, zhdankin_20}. Hence, we argue that orphan gamma-ray flares can be a self-consistent by-product of particle energization in magnetically-dominated plasmas with fast cooling.

The paper is organized as follows. In Section \ref{sec:model}, we present our model. In Section \ref{sec:simulations}, we validate our model using radiative fully-kinetic simulations of turbulent magnetically-dominated pair plasmas. Finally, in Section \ref{sec:conclusions} we conclude.

\section{Physical model}
\label{sec:model}

Let us consider a blob filled with a turbulent plasma. In the proper frame of the blob, the characteristic physical parameters of the plasma are (i) $l$, the turbulence energy-carrying scale, which we assume to be comparable to the size of the emitting blob; (ii) $B$, the intensity of the magnetic field. We assume the turbulent component of the magnetic field to be $\delta B\sim B$; (iii) $n_0$, the electron number density. The available magnetic energy per electron exceeds the rest mass, i.e.~the initial plasma magnetization is $\sigma_0\equiv \delta B^2/4\pi n_0 mc^2\gg 1$, where $m$ is the electron mass. If the plasma has a significant proton component, we assume that a large fraction (i.e., $\gtrsim 50\%$) of the turbulent energy heats the plasma electrons

The emitting blob moves with a velocity $\beta c$ and a Lorentz factor $\Gamma\equiv1/\sqrt{1-\beta^2}$ with respect to the observer. If the viewing angle is $\theta\lesssim 1/\Gamma$, the Doppler factor of the blob is $\delta\equiv 1/[\Gamma(1-\beta\cos\theta)] \sim\Gamma$. Throughout this paper we take a fiducial value of $\delta\sim\Gamma\sim 20$. For the physical parameters of blazars, such Lorentz factor typically guarantees that the gamma-rays do not annihilate with lower energy photons within the emitting blob \citep[e.g.][]{DondiGhisellini1995}.

The blob is immersed in an external radiation field of frequency $\nu_0\sim 10^{15}{\rm\; Hz}$ and energy density $U_{\rm rad}\sim 0.01{\rm\; erg\; cm}^{-3}$ (which in the blob frame are Lorentz-boosted to $\nu_0'\sim\Gamma\nu_0$, and $U'_{\rm rad}\sim\Gamma^2 U_{\rm rad}$), as may be produced by the Broad Line Region in FSRQ \citep[e.g.][]{Sikora1994, Sikora2009, GhiselliniTavecchio2009}. Due to IC losses, within a dynamical time $t_{\rm dyn}$ particles cool down to a Lorentz factor $\gamma_{\rm cool}\sim \max\left[1, 3mc/4\sigma_{\rm T}U'_{\rm rad}t_{\rm dyn}\right]$, where $\sigma_{\rm T}$ is the Thomson cross section.

The turbulent component of the magnetic field dissipates on a timescale $t_{\rm dyn}\sim l/c$ \citep[e.g.][]{ComissoSironi2018, ComissoSironi2019, NattilaBeloborodov2020}. We assume that (i) reconnection injects particles of Lorentz factor $\gamma\sim\sigma_0$ at a constant rate over a dynamical time, until the turbulent energy is dissipated; (ii) reconnection-accelerated particles cool efficiently within a dynamical time by IC scattering the external radiation field (i.e., as discussed in Section \ref{sec:discussion}, we consider the fast cooling regime $\gamma_{\rm cool} \ll \sigma_0$), which produces a gamma-ray flare. If the magnetic field within the emitting blob is tangled, the particle distribution is isotropic on the global scale of the blob (for example, the emitting blob may contain multiple turbulent cells, each having a random direction of the mean field). Then the IC emission is the same as in the standard isotropic case. In contrast, as we discuss below, the local particle distribution is strongly anisotropic in pitch angle. The anisotropy heavily suppresses the synchrotron emission.

We estimate the properties of the emitting plasma by considering a typical flare with isotropic equivalent luminosity $L_{\rm obs}\sim 10^{48}{\rm\; erg \; s}^{-1}$, duration $t_{\rm obs}\sim 1{\rm\; day}$, and peaking at a frequency $\nu_{\rm obs}\sim 10^{23}{\rm\; Hz}$ \citep[e.g.][]{Bottcher2019}.\footnote{Orphan flares with a variability timescale of a few hours were recently reported by \citet[][]{Patel+2020}. Variability on short timescales (i.e., $\ll 1{\rm\; day}$) may be produced by plasma clumps moving relativistically within the emitting region, as in the so-called ``jet-in-a-jet'' models \citep[e.g.][]{Giannios2009, Giannios2013}.} The duration of the flare in the observer's frame is $t_{\rm obs}\sim t_{\rm dyn}/\Gamma\sim l/\Gamma c$, which gives
\begin{equation}
\label{eq:l}
l\sim 5\times 10^{16} \left(\frac{\Gamma}{20}\right) \left(\frac{t_{\rm obs}}{1 {\rm\; day}}\right) {\rm\; cm} \;.
\end{equation}
The corresponding distance from the central engine may be estimated as $r\sim\Gamma l\sim 10^{18}{\rm\; cm}$. The isotropic equivalent of the flare luminosity is $L_{\rm obs}\sim (\delta B^2/8\pi)l^3\Gamma^4/t_{\rm dyn}$, which gives
\begin{equation}
\label{eq:B}
\delta B\sim 1.4 \left(\frac{\Gamma}{20}\right)^{-3} \left(\frac{t_{\rm obs}}{1{\rm\; day}}\right)^{-1} \left(\frac{L_{\rm obs}}{10^{48}{\rm\; erg\; s}^{-1}}\right)^{1/2} {\rm\; G}\;.
\end{equation}
Since IC scattering occurs in the Thomson regime, the peak frequency of the IC radiation is $\nu_{\rm obs}\sim \Gamma\sigma_0^2\nu_0'\sim \Gamma^2\sigma_0^2\nu_0$, which gives a magnetization\footnote{We remark that we have defined the magnetization with respect to the electron rest mass energy density.}
\begin{equation}
\label{eq:sigma_0}
\sigma_0\sim 500\left(\frac{\Gamma}{20}\right)^{-1} \left(\frac{\nu_{\rm obs}}{10^{23}{\rm\; Hz}}\right)^{1/2} \left(\frac{\nu_0}{10^{15}{\rm\; Hz}}\right)^{-1/2} \;.
\end{equation}
The IC emission extends down to a frequency $\nu\sim\Gamma^2\gamma_{\rm cool}^2\nu_0\ll\nu_{\rm obs}$, with a typical spectrum $\nu F_\nu\propto\nu^{1/2}$. From Eqs. \eqref{eq:B} and \eqref{eq:sigma_0}, one may estimate the particle number density, $n_0\sim \delta B^2/4\pi\sigma_0 mc^2\sim 10^2 {\rm\; cm}^{-3}$.

 \begin{figure}
\centering
    \includegraphics[width=0.49\textwidth]{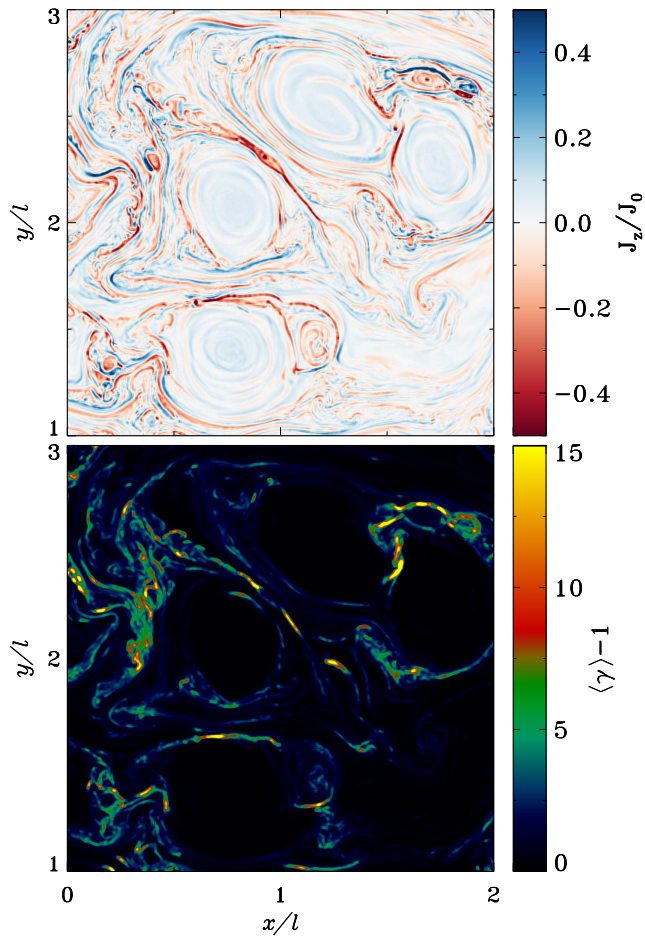}
\caption{
2D structure of turbulence from a simulation with $\sigma_0=160$, $\delta B_{\rm rms0}/B_0=0.5$, and $L/d_{e0}=13,300$
 (with $l=L/4$), at time $ct/l=7$ (near the peak of the lightcurves, see Figure \ref{fig:spectime}). We focus on a portion of the simulation domain to emphasize small-scale structures (the overall box is $4l\times 4l$). Top: out-of-plane current density $J_z$ (normalized to $J_0=e n_0 c$) indicating the presence of current sheets and reconnection plasmoids. Bottom: mean kinetic energy per particle (in units of $mc^2$), showing that high-energy particles are preferentially located at reconnecting current sheets.}
     \label{fig:fluid}
\end{figure}

\subsection{Effect of the anisotropy}
\label{sec:discussion}

Uncooled PIC simulations of magnetically-dominated pair plasma turbulence \citep[e.g.][]{ComissoSironi2018, ComissoSironi2019, Nattila2019, Wong2020} have shown that (i) particles are injected at Lorentz factors $\gamma\sim\sigma_0$ due to strong  non-ideal electric fields ($E_{\rm rec}\sim \eta_{\rm rec} \delta B$, with $\eta_{\rm rec}\sim 0.1$) in large-scale reconnecting current sheets. The injection time is $t_{\rm inj}\sim \sigma_0 mc/eE_{\rm rec}$, and typically $t_{\rm inj}\ll t_{\rm dyn}$. Since the reconnection electric field is nearly aligned with the local magnetic field, reconnection-accelerated particles have a strong anisotropy; (ii) particles may be further accelerated (up to $\gamma\gg\sigma_0$) by scattering off turbulent magnetic fluctuations. Scattering operates on a longer timescale, $t_{\rm scat}\sim t_{\rm dyn}$.

For the properties of the emitting plasma estimated above, one finds that $\gamma_{\rm cool}\sim 3mc/4\sigma_{\rm T}U'_{\rm rad}t_{\rm dyn}\sim 4$. The ratio of the injection to the cooling time for particles with $\gamma\sim\sigma_0$ is $t_{\rm inj}/t_{\rm cool}\sim (\sigma_0/\gamma_{\rm cr})^2\sim 10^{-8}$, where $\gamma_{\rm cr}$ is defined by the condition that IC losses balance the acceleration by the reconnection electric field, i.e.~$eE_{\rm rec}\sim 4\sigma_{\rm T}\gamma_{\rm cr}^2U'_{\rm rad}/3$. Since $t_{\rm inj}\ll t_{\rm cool}$, particle injection in large-scale current sheet is practically unaffected by cooling \citep[e.g.][]{NattilaBeloborodov2020, SobacchiLyubarsky2020}. The ratio of the cooling to the scattering time is $t_{\rm cool}/t_{\rm scat}\sim t_{\rm cool}/t_{\rm dyn}\sim \gamma_{\rm cool}/\sigma_0 \sim 10^{-2}$. Since $t_{\rm cool}\ll t_{\rm scat}$, further energization due to scattering is inhibited \citep[e.g.][]{NattilaBeloborodov2020, sironi_belo_20, SobacchiLyubarsky2020, zhdankin_20}. Effectively, in a fast cooling system, turbulence is only able to inject particles with $\gamma\sim\sigma_0$ via reconnection. Particles move nearly along the local magnetic field, and cool down to $\gamma \sim \gamma_{\mathrm{cool}}$ via IC scattering within a dynamical time. Then most of the particle energy is converted into radiation.

In the frame of the emitting blob, the magnetic energy density is $U_{\rm B}\equiv \delta B^2/8\pi\sim 0.08{\rm\; erg\; cm}^{-3}$, and the radiation energy density is $U'_{\rm rad}\sim 4{\rm\; erg\; cm}^{-3}$. If the particle distribution were isotropic, the synchrotron luminosity would be $L_{\rm sync}\sim (U_{\rm B}/U'_{\rm rad})L_{\rm obs}\sim 2\times 10^{46}{\rm\; erg\; s}^{-1}$. However, the synchrotron luminosity is suppressed by a factor $\sin^2\bar{\alpha}$, where $\bar{\alpha}\ll 1$ is the pitch angle between the particle velocity and the local magnetic field \citep[e.g.][]{ComissoSironi2019, comisso_sobacchi_20}. In the next section we demonstrate that this effect can suppress the synchrotron luminosity by orders of magnitude, making the low-energy counterpart of the gamma-ray flare practically undetectable.

Since we are considering IC scattering off an external radiation field, our model adequately describes orphan gamma-ray flares from FSRQ. In the case of BL Lacs, the seed photons are instead produced within the jet itself via synchrotron emission \citep[e.g.][]{Maraschi1992}. We argue that our proposed mechanism can explain orphan gamma-ray flares also in BL Lacs. Let us consider a fast cooling system, and neglect Klein-Nishina effects on IC scattering. The ratio of the synchrotron to IC power is $P_{\rm sync}/P_{\rm IC}\sim (U_{\rm B}/U_{\rm sync})\sin^2\bar{\alpha}$, where $U_{\rm sync}$ is the comoving energy density of the synchrotron radiation. Since only a fraction $P_{\rm sync}/P_{\rm IC}\ll 1$ of the available magnetic energy is converted to synchrotron radiation when $\bar{\alpha}\ll 1$, one finds that $U_{\rm sync}\sim (P_{\rm sync}/P_{\rm IC})U_{\rm B}$, and therefore $P_{\rm sync}/P_{\rm IC}\sim \sin\bar{\alpha}\ll 1$. Numerical simulations of this process are more challenging since the radiation field should be modeled self-consistently, and are left for future work.

\section{Particle-In-Cell Simulations}
\label{sec:simulations}

\subsection{Numerical method and setup}

We perform \emph{ab initio} PIC simulations employing the PIC codes TRISTAN-MP \citep{buneman_93,spitkovsky_05} and \textsc{Runko} \citep{Nattila2019}. Simulations were performed with both the codes, finding consistent results. We conduct large-scale two-dimensional (2D) simulations in the $xy$ plane, but all three components of particle momenta and electromagnetic fields are evolved in time.\footnote{Our previous studies have demonstrated that the particle energy distribution and anisotropy are nearly the same between 2D and 3D for uncooled systems \citep{ComissoSironi2018, ComissoSironi2019, comisso_sobacchi_20}. In uncooled systems, particles are first energized by reconnection, and then by scattering off turbulent magnetic fluctuations. Energization due to scattering is arguably sensitive to 3D effects since turbulent fluctuations are anisotropic. Then we expect 3D effects to be minor in fast cooling systems, since energization due to scattering becomes ineffective.} The computational domain is a square of size $L\times L$, with periodic boundary conditions in all directions.

\begin{figure}
\centering
    \includegraphics[width=0.49\textwidth]{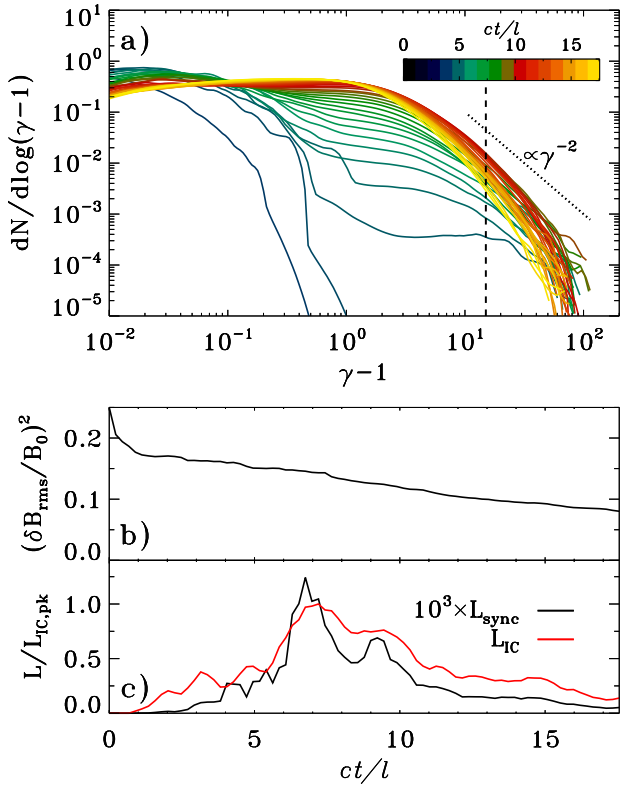}
\caption{Temporal evolution of the same simulation as in Figure \ref{fig:fluid}. We show: (a) the time evolution of the particle energy spectrum; the dotted line indicates the slope $dN/d\log(\gamma-1)\propto \gamma^{-2}$ yielding equal IC luminosity per decade in Lorentz factor. (b) the decay of the magnetic energy density in turbulent fluctuations. (c) IC (red) and synchrotron (black) lightcurves, normalized to the peak of the IC curve;  the synchrotron emission is scaled up by a factor of $10^3$. Both lightcurves are computed using particles with Lorentz factor $\gamma\geq16=\sigma_0/10$ (to the right of the dashed line in panel (a)).
}
\label{fig:spectime}
\end{figure}

The simulation setup is similar to our previous works on magnetically-dominated plasma turbulence \citep{ComissoSironi2018, ComissoSironi2019, comisso_sobacchi_20, NattilaBeloborodov2020}. We initialize a uniform electron-positron plasma with total particle density $n_0$ and a small thermal spread $\theta_0 = {k_{\rm B} T_0}/{m c^2} = 10^{-5}$, where $k_{\rm B}$ is the Boltzmann constant and $T_0$ is the initial plasma temperature (we have obtained practically identical results using a thermal spread $\theta_0=0.1$, in which case the Debye length is resolved since the very beginning of the simulation). Turbulence develops from uncorrelated magnetic field fluctuations that are initialized in the plane perpendicular to a uniform mean magnetic field $B_0 {\bf e}_z$. The initial fluctuations have low wavenumbers $k_j = 2\pi n_j/L$, where $n_j \in \{ {1, \ldots ,4} \}$ and $j$ indicates the wavenumber direction, and equal amplitude per mode. The initial magnetic energy spectrum peaks near $k_p = 8 \pi /L$. We use $l = 2 \pi /k_p=L/4$ as our unit length.

The strength of the initial magnetic field fluctuations is parameterized by the magnetization $\sigma_0  = \delta B_{{\rm{rms}}0}^2/4\pi n_0 m c^2$, where  $\delta B_{{\rm{rms}}0} = \langle {\delta {B^2} (t=0)} \rangle^{1/2}$ is the space-averaged root-mean-square value of the initial fluctuating fields. We use a fiducial magnetization $\sigma_0=160$, but we have obtained similar results for  $\sigma_0\gtrsim 40$. We vary the level of turbulent fluctuations $\delta B_{\rm rms0}/B_0$ between 0.5 (our fiducial case) and 2. This plays a significant role on the pitch angle distribution of accelerated particles, and so on the resulting synchrotron emission.

The large size of our computational domain (with $L=40,000$ cells) allows to achieve asymptotically-converged results. We resolve the initial plasma skin depth $d_{e0}=c/\omega_{\rm p0}=\sqrt {{m}c^2/4\pi n_0 {e^2}}$ with 3 cells, and we employ 4 computational particles per cell. The simulation time step is controlled by the numerical speed of light of 0.45 cells per time step. Our earlier studies \citep{ComissoSironi2018,ComissoSironi2019} have demonstrated convergence with respect to these numerical parameters.

We implement IC cooling as a Compton drag term in the particle equation of motion \citep[e.g.][]{werner_19, NattilaBeloborodov2020, sironi_belo_20, zhdankin_20}, assuming that the radiation is isotropic in the simulation frame and that Compton scattering happens in the Thomson regime. We parameterize the strength of IC cooling by defining the Lorentz factor $\gamma_{\rm cr}$ at which IC losses would prohibit further acceleration by the reconnection electric field $E_{\rm rec} = \eta_{\rm rec} \delta B_{\rm rms 0}$ ($\eta_{\rm rec}\sim 0.1$), i.e.~$eE_{\rm rec}=4\sigma_{\rm T}\gamma_{\rm cr}^2U'_{\rm rad}/3$. We choose a reference value of $\gamma_{\rm cr}=80$, which is larger than the mean Lorentz factor $\sim \sigma_0/4=40$ at which particles are accelerated by reconnection. The Lorentz factor of the cooled particles may be written as $\gamma_{\rm cool}\sim\max [1, (\gamma_{\rm cr}^2 / \eta_{\rm rec} \sqrt{\sigma_0})(d_{e0} / ct_{\rm dyn})]$, where $t_{\rm dyn}\sim l/c$. For our reference choice of $l=L/4\simeq 3,300\,{d_{e0}}$, one finds that $\gamma_{\rm cool}\sim 1.5$.

Our simulations maintain the correct ordering of the relevant timescales ($t_{\rm inj}<t_{\rm cool}<t_{\rm scat}\sim t_{\rm dyn}$), as it is expected for blazar conditions (see Section \ref{sec:discussion}). However, reproducing the correct separation between these timescales would require an impossibly large simulation box. Then our numerical results demonstrate that IC emission may have a negligible synchrotron counterpart as an effect of small pitch angles, but are not meant to reproduce the exact properties (e.g., peak frequency and luminosity) of blazar flares.

\begin{figure}
\centering
 \includegraphics[width=0.49\textwidth]{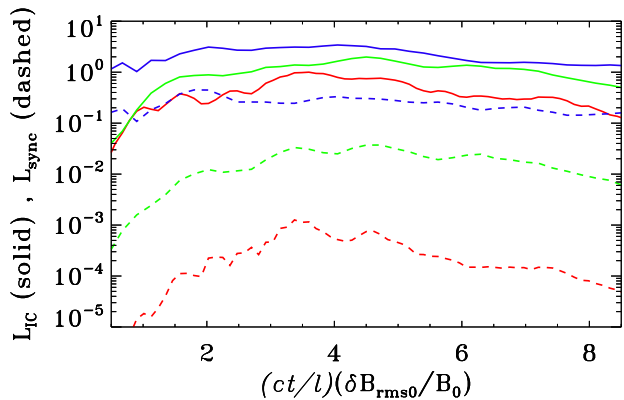}
\caption{Dependence of the IC (solid) and synchrotron (dashed) lightcurves on the level of turbulent fluctuations $\delta B_{\rm rms0}/B_0$, for our reference case $\delta B_{\rm rms0}/B_0=0.5$ (red) and two cases with stronger fluctuations: $\delta B_{\rm rms0}/B_0=1$ (green) and $\delta B_{\rm rms0}/B_0=2$ (blue). Time on the horizontal axis is rescaled with $\delta B_{\rm rms0}/B_0$ to account for the different rate of turbulence decay.
}
    \label{fig:lccomp}
\end{figure}

\subsection{Results}

In our simulations, turbulence develops from the initial unbalanced state, and the magnetic energy decays over time since we do not impose any continuous external driving. As shown in Figure \ref{fig:fluid} (top panel), the turbulent cascade leads to the formation of intense current layers. Several layers become prone to fast magnetic reconnection due to the plasmoid instability \citep[e.g.][]{comisso_16, uzdensky_16}. Magnetic reconnection plays a crucial role in extracting particles from the thermal pool and injecting them into the acceleration process \citep[e.g.][]{ComissoSironi2018,ComissoSironi2019}. In the absence of cooling, particles energized by reconnection would be accelerated to even higher energies by scattering off the turbulent fluctuations. However, the rate of stochastic acceleration is slower than the IC cooling rate, so particle acceleration to $\gamma\gg\sigma_0$ is inhibited. Rather, at any given time the highest energy particles in the simulation are still experiencing significant reconnection-powered acceleration. In fact, the locations of  high-energy particles (see the mean kinetic energy per particle in the bottom panel of Figure \ref{fig:fluid}) are well correlated with reconnection layers (compare top and bottom panels).
  
  In Figure \ref{fig:spectime}(a), we show the time evolution of the particle spectrum $dN/d\log(\gamma-1)$. As a result of field dissipation, the spectrum shifts to energies much larger than the initial thermal energy. At late times the spectral peak is at $\gamma\sim 2$, and it is populated by cooled particles (in fact, $\gamma_{\rm cool}\sim 1$ for our simulation). The highest energy part of the spectrum (at $\gamma\gtrsim 16$, beyond the vertical dashed line) builds up at early times, reaches the highest normalization at $5\lesssim ct/l\lesssim 10$, and is depleted at later times. The depletion at late times is due to the fact that in our simulation the magnetic energy decays over time (Figure \ref{fig:spectime}(b)), so turbulence-induced reconnection progressively becomes less efficient. The resulting decrease in magnetization is associated with an analogous drop in the high-energy spectral cutoff of the Lorentz factor of reconnection-accelerated particles, whose Lorentz factor is $\gamma\sim\sigma_0/4=40$, as seen in Figure \ref{fig:spectime}(a).
  
In order to compute the IC and synchrotron lightcurves, we only consider particles with Lorentz factors $\gamma\geq16=\sigma_0/10$ (i.e., to the right of the vertical dashed line in Figure \ref{fig:spectime}(a)), which dominate the emission. For each particle, we compute the IC power as $P_{\rm IC}\propto \gamma^2 U'_{\rm rad}$, whereas the synchrotron power is $P_{\rm sync}\propto \bar{\gamma}^2\bar{U}_B\,\sin^2\bar{\alpha}$. For the latter, the particle Lorentz factor $\bar{\gamma}$, the magnetic energy density $\bar{U}_B$ and the pitch angle $\bar{\alpha}$ are all computed in the local $E\times B$ frame (i.e., the frame moving with the local $E\times B$ speed). The resulting lightcurves are presented in Figure \ref{fig:spectime}(c), assuming equipartition of radiation and magnetic energy density in the simulation frame at the initial time (i.e., $U'_{\rm rad}=U_{\rm B0}=B_0^2/8 \pi$; note that the magnetic energy is dominated by the mean field for our reference case with $\delta B_{\rm rms0}/B_0=0.5$). Both lightcurves peak at $ct/l\sim 7$, and display significant emission in the time interval $5\lesssim ct/l \lesssim 10$. However, as indicated in the legend, the synchrotron emission (black line) is suppressed by three orders of magnitude, as compared to the IC emission (red curve). This follows from the small pitch angles of  particles accelerated by reconnection. In fact, given that we assume $U'_{\rm rad}=U_{\rm B0}$, the ratio of synchrotron to IC emission is expected to be approximately equal to the mean value of $\sin^2\bar{\alpha}$, which we indeed measure to be $\sim 10^{-3}$. In summary, the small pitch angle of reconnection-accelerated particles is responsible for a significant suppression of the synchrotron luminosity.
  
In Figure \ref{fig:lccomp}, we present the dependence of the IC (solid) and synchrotron (dashed) lightcurves on the level of turbulent fluctuations $\delta B_{\rm rms0}/B_0$, by comparing our reference case $\delta B_{\rm rms0}/B_0=0.5$ (red) with two cases with stronger fluctuations: $\delta B_{\rm rms0}/B_0=1$ (green) and $\delta B_{\rm rms0}/B_0=2$ (blue). All curves are normalized to the peak IC luminosity of our reference case and they assume $U'_{\rm rad}=U_{\rm B0}$. The figure shows that the ratio of IC to synchrotron luminosity is a strong function of $\delta B_{\rm rms0}/B_0$. For larger fluctuations, the particles accelerated by reconnection have larger pitch angles (as also found in the uncooled simulations of \citealt{comisso_sobacchi_20}), and so they are more efficient synchrotron emitters. In fact, the synchrotron luminosity increases by nearly two orders of magnitude from $\delta B_{\rm rms0}/B_0=0.5$ to $\delta B_{\rm rms0}/B_0=1$, whereas the IC luminosity is nearly unchanged. The modest increase in IC luminosity with $\delta B_{\rm rms0}/B_0$ is in agreement with results of uncooled simulations, where higher $\delta B_{\rm rms0}/B_0$ yielded harder particle spectra, and so more efficient particle acceleration to high energies \citep[e.g.][]{ComissoSironi2018}.
 
 We have checked that the anisotropy and energy spectrum of high-energy particles are independent of the box size if $\sigma_0$ and $\gamma_{\rm cr}^2 (d_{e0}/l)$ are kept constant. Since also $\gamma_{\rm cool}\sim\max [1, (\gamma_{\rm cr}^2 / \eta_{\rm rec} \sqrt{\sigma_0})(d_{e0} / ct_{\rm dyn})]$ remains constant, our results are applicable to asymptotically large astrophysical systems in the fast cooling regime.

\section{Conclusions}
\label{sec:conclusions}

In this work, we show that orphan gamma-ray flares can be a self-consistent by-product of the particle acceleration physics in magnetically-dominated pair plasmas. We mostly focus on orphan GeV flares from Flat Spectrum Radio Quasars. Particles energized by the decaying turbulence cool down on a dynamical time by IC scattering an external photon field (e.g.~the photons emitted by the Broad Line Region), thus producing the gamma-ray flare. The flare has a faint low-energy counterpart since the particles are accelerated nearly along the local magnetic field, and then the synchrotron emission is suppressed.

Even though our numerical results are based on simulations of magnetically-dominated plasma turbulence, we argue that our conclusions hold more generally, for any system where particle injection is governed by reconnection in the strong guide field regime, and where fast cooling prevents further particle energization. This may happen in the non-linear stages of, e.g., the kink instability \citep{davelaar_20} and the Kelvin-Helmholtz instability \citep{sironi_20}.

The majority of blazar gamma-ray flares have a luminous low-energy counterpart. Our results show that the ratio of inverse Compton to synchrotron luminosity may be regulated by the initial strength of the turbulent fluctuations (the anisotropy of the accelerated particles is weaker for a larger degree of fluctuations, and the synchrotron luminosity increases). A complementary possibility, yet to be tested with PIC simulations, is that kinetic instabilities reduce the anisotropy if the rest mass energy density of the plasma is dominated by the ions \citep{SobacchiLyubarsky2019}.

Our numerical results rely on 2D simulations. Our previous studies have demonstrated that the particle energy distribution and anisotropy are nearly the same between 2D and 3D for uncooled systems \citep{ComissoSironi2018, ComissoSironi2019, comisso_sobacchi_20}. Investigating 3D effects in fast cooling systems is a worthwhile direction for future investigation.

\section*{Acknowledgements}

We thank the anonymous referee for constructive comments and suggestions that improved the paper. We are grateful to Luca Comisso for insightful discussions. LS acknowledges support from the Sloan Fellowship, the Cottrell Scholar Award, DoE DE-SC0021254,  NASA ATP 80NSSC18K1104, and NSF PHY-1903412. The simulations have been performed at Columbia (Habanero and Terremoto), and with NERSC (Cori) and NASA (Pleiades) resources.

\section*{Data availability}

The data underlying this article will be shared on reasonable request to the corresponding author.

\def\aap{A\&A}\def\aj{AJ}\def\apj{ApJ}\def\apjl{ApJ}\def\mnras{MNRAS}\def\prl{Physical Review Letters}
\def\araa{ARA\&A}\def\physrep{PhR}\def\sovast{Sov. Astron.}\def\nar{NewAR}
\def\aapr{Astronomy \& Astrophysics Review}\def\apjs{ApJS}\def\nat{Nature}\def\na{New Astron.}
\def\prd{Phys. Rev. D}\def\pre{Phys. Rev. E}\def\pasp{PASP}\def\ssr{Space Sci. Rev.}
\bibliographystyle{mn2e}
\bibliography{2d}

\end{document}